# Effect of atmospheric turbulence on propagation properties of optical vortices formed by using coherent laser beam arrays

Li-Gang Wang,[1, 2] [*] Wei-Wei Zheng,[1] and Li-Qin Wang[1]

[1] *Department of Physics, Zhejiang University, Hangzhou 310027, China*
[2] *Department of Physics, The Chinese University of Hong Kong, Shatin, N. T., Hong Kong, China*
[*]*Corresponding author: sxwlg@yahoo.com.cn*

**Abstract:** In this paper, we consider the effect of the atmospheric turbulence on the propagation of optical vertex formed from the radial coherent laser beam array, with the initially well-defined phase distribution. The propagation formula of the radial coherent laser array passing through the turbulent atmosphere is analytically derived by using the extended Huygens-Fresnel diffraction integral. Based on the derived formula, the effect of the atmospheric turbulence on the propagation properties of such laser arrays has been studied in great detail. Our main results show that the atmospheric turbulence may result in the prohibition of the formation of the optical vortex or the disappearance of the formed optical vortex, which are very different from that in the free space. The formed optical vortex with the higher topological charge may propagate over a much longer distance in the moderate or weak turbulent atmosphere. After the sufficient long-distance atmospheric propagation, all the output beams (even with initially different phase distributions) finally lose the vortex property and gradually become the Gaussian-shaped beams, and in this case the output beams actually become incoherent light fields due to the decoherence effect of the turbulent atmosphere.

©2008 Optical Society of America

## 1. Introduction

Optical vortices with the topological wave-front dislocations have attracted intensive attentions in many branches of classical and quantum physics [1-3], due to their important applications, such as optical testing [4], optical tweezers [5-6], quantum entanglement [3, 7, 8], and stellar coronagraph [9, 10], etc. Nowadays, optical vortices can be generated by various methods such as mode conversions [11-12], computer generated holograms [13], spiral phase plates [14-15] and multi-level spiral phase plate [16], and optical wedges [17, 18]. Recently, the interference of several plane waves was used to generate the optical vortices [19]. The modified Michelson interferometer and the modified Mach-Zehnder interferometer were presented for producing optical vortex arrays [20]. The stable rotating structures in the optical lattices have been formed, using the rotating counterpropagating incoherent self-trapped vortex beams [21]. The singular beams with high orbital angular momentum could be experimentally obtained by using an array of vortex beams [22]. In our recent work, we have used a radial beam array composed of coherent Gaussian beamlets to generate optical vortices. In that work, we arranged the fundamental Gaussian beamlets with the initial well-ordered phase in a radial symmetric configuration, and found the stable optical vortices formed at the far-field region in the free space [23].

Although there are many investigations on the propagation properties of various coherent and partially coherent laser beams in a turbulent atmosphere [24-33], the effect of atmospheric turbulence on the propagation of the optical vortex is still few studied. In the previous studies on the atmospheric propagation, the spreading effect of the laser beam in the turbulent atmosphere has been intensively investigated. In the singular optics [34], one important parameter is the topological charge (a measure of the angular momentum of the beam). Recently, Vinçotte and Bergéhave examined the robustness of the femtosecond optical vortices propagating freely in the atmosphere and have demonstrated that the pulsed optical vortices can propagate over hundreds of meters before they break up into filaments. [35] Gbur and Tyson have showed that the topological charge of the optical vortex is a robust quantity that can be transmitted over a significant distance without loss. [36] In this paper, following our recent work [23], we would like to further investigate the effect of the turbulent atmosphere on the propagation of the optical vortex formed from a radial coherent laser beam array, with the initially well-ordered phase distribution. We show that the atmospheric turbulence may result in the prohibition of the formation of the optical vortex or the disappearance of the formed vortex properties after a sufficient long-distance propagation.

## 2. Theoretical model of the radial coherent laser beam array and its propagation in the turbulent atmosphere

It is well known that the laser arrays have been investigated widely because of their potential applications in high-energy weapon and atmospheric optical communication [37-41]. In all the previous studies [37-42], the beam arrays in the radial or rectangular symmetrical configurations are investigated in both phase-locked and non-phase-locked cases. For the phase-locked case it usually refers to the coherent superposition of all beamlets with the same initial phases, while for the non-phase locked case it refers to the incoherent addition of all beamlet [40-44]. However, unlike the previous phase-locked radial laser arrays, each beamlet of the radial coherent beam arrays considered in our recent work [23] has the different initial phase in a well-ordered distribution. It has been shown that the optical vortices could be formed during the propagation of such radial symmetric beam arrays in free space.

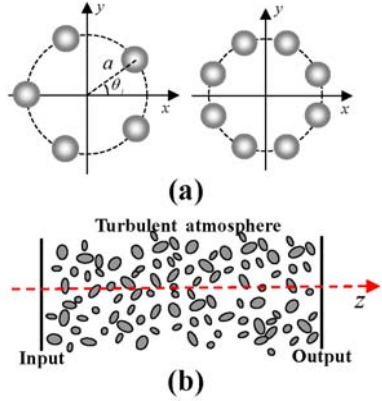

FIG. 1 (a) Schematic of a radial laser arrays with $N=5,8$ identical fundamental Gaussian beams, which are uniformly located on a ring with radius $a$ at the azimuth angle $\theta_j$ ( $j=1,\cdots,N$ ). (b) Schematic of a turbulent atmospheric optical system.

Following our recent work [23], we consider the radial laser array consisted of $N$ identical Gaussian beams, which are located symmetrically on a ring with radius $a$ (the dashed circle), as shown in Fig.1(a). Every beamlet has the same half waist-width $w_0$ and different initial phases $\varphi_j$, and each beamlet's center coordinates $(\alpha_j,\beta_j)$ are $(a\cos\theta_j, a\sin\theta_j)$, where $\theta_j = \pi(2j-1)/N$ is the azimuth angle of the $j$ th beamlet and the radius $a$ controls the separation distance among these beamlets under the fixed number $N$. As we have introduced in our previous work [23], when the initial phase $\varphi_j$ of each beamlet is well-ordered distributed in the following way

$$\varphi_j = \pi m(2j-1)/N, \tag{1}$$

where $m$ is related to the topological charge of the resulted beam, one always can obtain the optical vortex at the far-field region in the vacuum or near the focus plane in the lens focusing system. [23]

Now consider such a radial laser array propagating through a turbulent atmosphere, as shown in Fig. 1(b). For the $j$ th initial Gaussian beamlet at the input plane $z=0$ (source plane), its field distribution in the rectangular coordinates is defined by

$$E_j(x_1,y_1;0) = G_0 e^{-\frac{(x_1-\alpha_j)^2+(y_1-\beta_j)^2}{2w_0^2}} e^{i\varphi_j}, \tag{2}$$

where $j=1,2,\cdots,N$ denotes the list of the beamlets, the initial phase $\varphi_j$ is given by Eq. (1), and $G_0$ is a constant. Therefore the total field of the radial laser array in the input plane ( $z=0$ ) can be expressed as

$$E_{in}(x_1,y_1;0) = \sum_{j=1}^{N} E_j(x_1,y_1;0) = G_0 \sum_{j=1}^{N} e^{-\frac{(x_1-\alpha_j)^2+(y_1-\beta_j)^2}{2w_0^2}} e^{i\varphi_j}. \tag{3}$$

The field distribution of the resulting beam in the output plane ( $z>0$ ) can be determined from the knowledge of the field distribution in the input plane ( $z=0$ ) by using the extended Huygens-Fresnel diffraction integral as follows: [45]

$$E_{out}(x_2, y_2; z) = -\frac{ik}{2\pi z} e^{ikz} \int_{-\infty}^{\infty} \int_{-\infty}^{\infty} E_{in}(x_1, y_1; 0) e^{\frac{ik}{2z}\left[(x_1-x_2)^2 + (y_1-y_2)^2\right] + \psi(x_1, y_1, x_2, y_2, z)} dx_1 dy_1, \quad (4)$$

where $k = 2\pi/\lambda$ is the wave number of the incident Gaussian beamlet with wavelength $\lambda$, $(x_2, y_2)$ are the coordinates in the output plane, and $\psi(x_1, y_1, x_2, y_2, z)$ represents the random factor in the Rytov method due to the turbulent atmosphere for the complex phase of a spherical wave propagating from the input plane to the output plane.[25-29] In the current situations, we are interested in the average intensity distribution at the output plane and the phase information of the output beam becomes meaningless due to the random factor $\psi(x_1, y_1, x_2, y_2, z)$.

On submitting from Eq. (3) into Eq. (4), the total average intensity distribution of the output beam can be written as

$$I_{out}(x_2, y_2; z) = \left\langle E_{out}(x_2, y_2; z) E_{out}^*(x_2, y_2; z) \right\rangle_e$$

$$= \frac{k^2 G_0^2}{4\pi^2 z^2} \int_{-\infty}^{\infty}\int_{-\infty}^{\infty}\int_{-\infty}^{\infty}\int_{-\infty}^{\infty} dx_1 dy_1 dx'_1 dy'_1 \sum_{j=1}^{N}\sum_{l=1}^{N} e^{-\frac{(x_1-\alpha_j)^2+(y_1-\beta_j)^2}{2w_0^2}} e^{-\frac{(x'_1-\alpha_l)^2+(y'_1-\beta_l)^2}{2w_0^2}} e^{i\varphi_j - i\varphi_l}$$

$$\times e^{\frac{ik}{2z}\left[(x_1-x_2)^2+(y_1-y_2)^2\right] - \frac{ik}{2z}\left[(x'_1-x_2)^2+(y'_1-y_2)^2\right]} \left\langle e^{\psi(x_1,y_1,x_2,y_2,z)+\psi^*(x'_1,y'_1,x_2,y_2,z)} \right\rangle_e.$$

(5)

where $\langle \rangle_e$ denotes the ensemble average. The statistical ensemble average over the random complex phase for the turbulence of the homogeneous atmosphere could be expressed as [46]

$$\left\langle e^{\psi(x_1,y_1,x_2,y_2,z)+\psi^*(x'_1,y'_1,x_2,y_2,z)} \right\rangle_e = e^{-\frac{1}{2}D_\psi(x_1,y_1;x'_1,y'_1)} \cong e^{-\frac{1}{\rho_0^2}[(x_1-x'_1)^2+(y_1-y'_1)^2]}, \quad (6)$$

where $D_\psi$ is the wave structure function of the random complex phase in Rytov's representation, and $\rho_0 = (0.545 C_n^2 k^2 z)^{-3/5}$ is the coherence length of a spherical wave propagating in a turbulent atmosphere that is characterized by the refractive index structure parameter $C_n^2$. In Eq. (6), we only keep the quadratic terms in the Rytov's phase structure function and the higher-order terms are neglected in order to obtain the analytical results and the direct physical pictures. It has been pointed out that such an approximation made in Eq. (6) is reasonable and valid not only for weak turbulence but also for strong turbulence [25].

On substituting Eq. (6) into Eq. (5) and using the Gaussian integral formula $\int \exp[-Cx^2] dx = \sqrt{\pi/C}$ ($\text{Re}[C] > 0$), after the tedious calculations, we finally obtain:

$$I_{out}(x_2, y_2; z) = \frac{G_0^2 w_0^2}{w^2(z)} \exp\left[-\frac{x_2^2 + y_2^2}{w^2(z)}\right] \exp\left[-\frac{a^2}{w^2(z)}\right]$$

$$\times \sum_{j=1}^{N}\sum_{l=1}^{N} e^{i(\varphi_j - \varphi_l)} e^{\frac{1}{w^2(z)}[a(\cos\theta_j + \cos\theta_l) - \frac{iza}{kw_0^2}(\cos\theta_j - \cos\theta_l)]x_2} \quad (7)$$

$$\times e^{\frac{1}{w^2(z)}[a(\sin\theta_j + \sin\theta_l) - \frac{iza}{kw_0^2}(\sin\theta_j - \sin\theta_l)]y_2} e^{-\frac{2a^2 z^2[1-\cos(\theta_j - \theta_l)]}{w^2(z) k^2 w_0^2 \rho_0^2}}.$$

In the above derivations, the equalities $\alpha_p^2 + \beta_p^2 = a^2$, $\alpha_p = a\cos\theta_p$ and $\beta_p = a\sin\theta_p$ (the subscript $p = j, l$) have been used. Here $w(z) = \left[w_0^2 + \frac{\lambda^2 z^2}{\pi^2}\left(\frac{1}{4w_0^2} + \frac{1}{\rho_0^2}\right)\right]^{1/2}$ is the change of the beam half-width for the single Gaussian beamlet in the radial laser beam array. Equation (7) provides a general description of the radial coherent beam arrays passing through a turbulent atmosphere. The main difference between the output beams from the vacuum and the turbulent atmosphere is the factor $\rho_0^2$ in Eq. (7), which strongly affects the interference between different beamlets and the coherent superposition of each beamlet in the radial laser beam array. In fact, from Eq. (6), one can see that the coherence length $\rho_0$ of the light field in the turbulent atmosphere decreases in the form of $z^{-3/5}$ as the propagating distance $z$ increases, and this fact leads to the output beam becomes a completely incoherent beam after a sufficient long-distance atmospheric propagation. For the stronger atmospheric turbulence (i.e., $C_n^2$ increases), the value of $\rho_0$ decreases much faster. Therefore, with the increasing of the propagating distance, the output beam will gradually erase the information on the initial phase distributions in the radial laser beam array and the vortex properties of the output beam gradually disappear. In the following discussions, we use Eq. (7) to study the propagation properties of such radial coherent beam arrays passing through the turbulent atmosphere. In the following simulations, we take the beam array's parameters as follows: $\lambda = 632.8$ nm and $w_0 = 1$ mm.

## 3. Numerical results and discussions

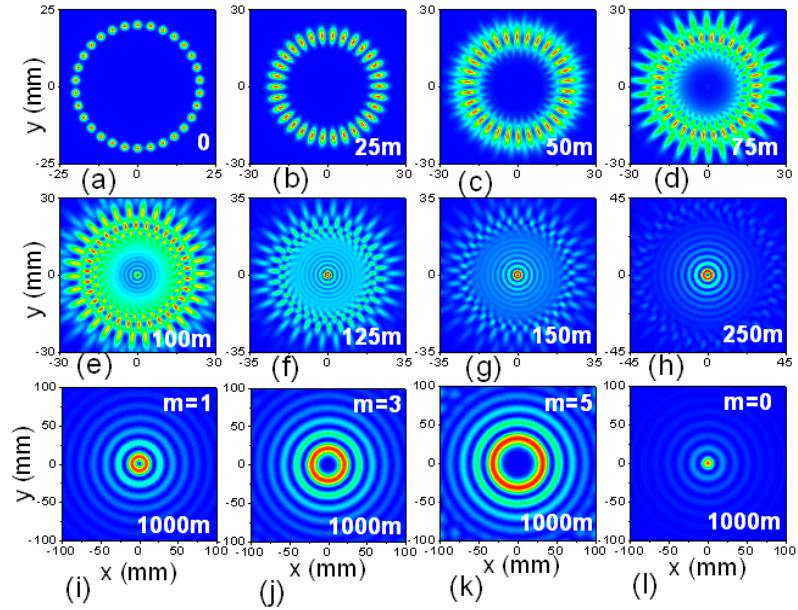

Fig. 2. (Color Online). (a)-(i) Evolution of the relative intensities of the radial beam array (with $m = 1$) at different output planes in the free space ($C_n^2 = 0$): (a) $z = 0$, (b) $z = 25$ m, (c) $z = 50$ m, (d) $z = 75$ m, (e) $z = 100$ m, (f) $z = 125$ m, (g) $z = 150$ m, (h) $z = 250$ m, and (i) $z = 1000$ m. For comparison, the relative intensities for the radial beam arrays with $m = 3$, 5 at the output plane $z = 1000$ m are, respectively, shown in (j) and (k), and the case for $m = 0$ is shown in (l). Other parameters are $N = 30$ and $a = 20w_0$.

Before we consider the effect of the turbulent atmosphere on the propagation properties of the radial coherent beam array, we first plot Fig. 2 to show the typical evolution of the relative intensity distribution for the radial beam array (with $N = 30$ and $m = 1$) propagating through the free spaces ($C_n^2 = 0$). As we have pointed out in Ref. [23] that the intensity distribution of the output beam is gradually rotating clockwise along the optical axis $z$ and the light energy also clockwise flow to form the inner annular beams [see Fig. 2(d) to 2(h)]. In fact, these inner annular beams in (h) and (i) are optical vortices with the topological charge $m$ [determined by the initial phase distribution of Eq. (1)], which could be easily judged from the phase distribution as pointed out in Ref. [23] (also see the detail in Ref. [23]). Figure 2(i) to 2(l) show the intensity distributions of the resulting beams at the output planes $z = 1$ km in free space when the initial radial laser beam arrays have the initial phase distributions with different $m = 1, 3, 5, 0$. Clearly, with the larger $m$, the inner multi-annular beams have the larger scales, while for the case of $m = 0$, the center of the result beam becomes bright and no optical vortex exists.

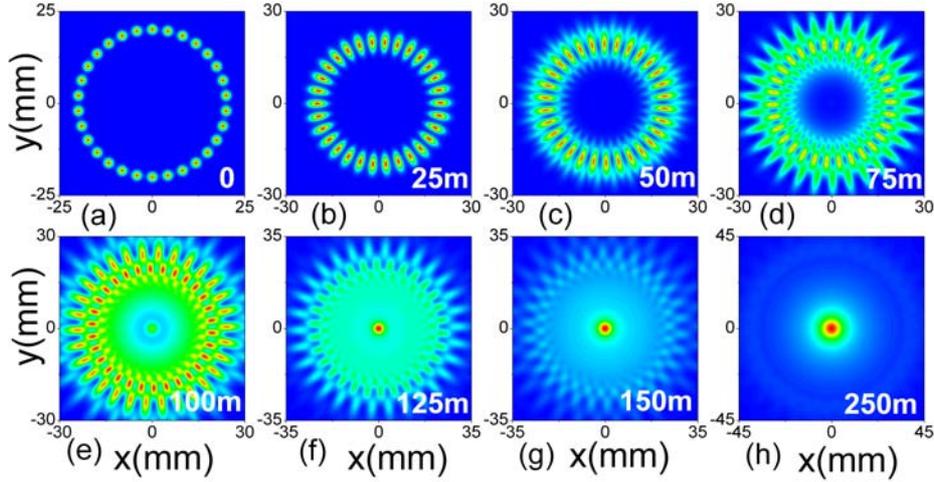

Fig. 3. (a)-(h) The relative intensities of the radial beam array passing through the strong turbulent atmosphere ($C_n^2 = 1 \times 10^{-13}$ m$^{-2/3}$) at different output planes: (a) $z = 0$, (b) $z = 25$ m, (c) $z = 50$ m and (d) $z = 75$ m, (e) $z = 100$ m, (f) $z = 125$ m, (g) $z = 150$ m and (h) $z = 250$ m, with other parameters $N = 30$, $m = 1$ and $a = 20w_0$.

Now we study the propagation properties of the radial coherent laser beam array, with the initial well-ordered phase distribution [see Eq. (1)], passing through a turbulent atmosphere by using the analytical formula derived in the previous section. Figure 3 shows the typical evolution of the relative average intensity of such a laser beam array passing through a strong turbulent atmosphere ($C_n^2 = 1 \times 10^{-13}$ m$^{-2/3}$). One can see from Fig. 3 that the evolution properties of the intensity distribution of the resulting beam in the turbulent atmosphere are similar to that in free space within the short-distance region ($z < 100$ m), namely, each beamlet of the radial beam array interferences each other and the intensity distribution of the output beam gradually rotates clockwise as the propagating distance $z$ increases. While as the propagating distance $z$ further increases, although the outside intensity distribution has the tendency to clockwise flow the energy into the center part, the coherence and superposition properties of the output beam is gradually destroyed due to the effect of the strong atmospheric turbulence (i.e., the coherence length $\rho_0$ in the turbulent atmosphere decreases in the form of $z^{-3/5}$.) Thus the resulting beam gradually loses all the initial phase information in the initial radial beam array and becomes a Gaussian-like beam after a sufficient long-distance propagation. This evolution process of the intensity distribution of the output beam in the atmosphere is totally different from that in the free space by comparing Fig. 3 with Fig. 2 (a-h). Therefore the strong atmospheric turbulence may strongly affect the formations of the multi-annular intensity distribution and the optical vortex of the output beams.

Figure 4 also shows the similar evolution process of the relative average intensity of the radial coherent laser beam array passing through a moderate turbulent atmosphere ($C_n^2 = 1 \times 10^{-15}$ m$^{-2/3}$). In this case, due to the fact that the atmospheric turbulence becomes weaker than that in Fig. 3, the optical vortex and homocentric multi-annular centric-dark beam could be formed in the short-distance region ($z < 1$ km) as similar to the case in the free space. However, as the propagating distance $z$ further increases, the vortex properties formed from the coherent superposition of the radial beam array will gradually disappear [see Figs. 4(c), 4(d) and 4(e)], due to the effect of the turbulent atmosphere, which strongly decreases the coherence of the output beam and destroys the coherent superposition process. From Fig. 4(e), after the sufficient long-distance propagation, one can see that the final output beam profile becomes a Gaussian-shaped beam again, similar to the case in Fig. 3(h). The physical explanation on the final output beam profile will be discussed in the below. Further, comparing the evolution in Fig. 4 with that in Fig. 3, one can find that the weaker the atmospheric turbulence (i.e., the smaller the structure parameter $C_n^2$), the longer distance the optical vortices could propagate over.

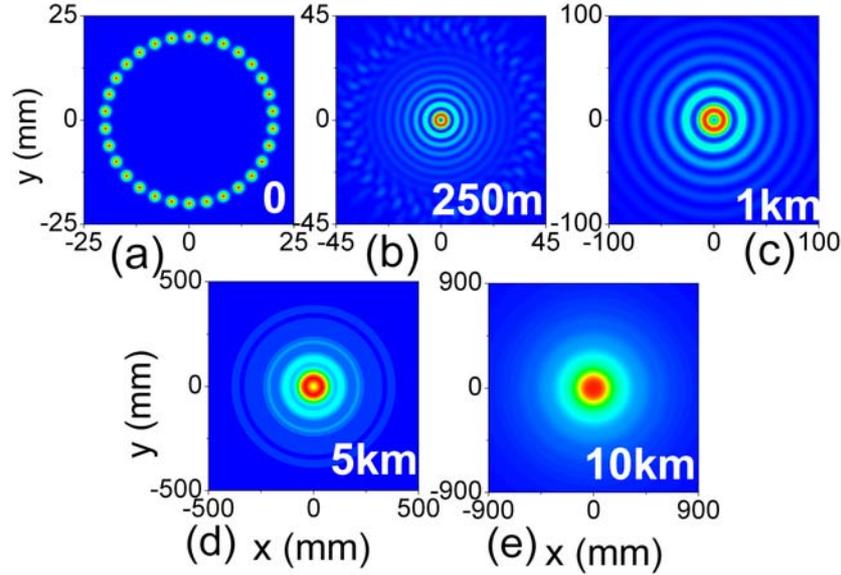

Fig. 4. (Color Online). (a)-(e) The relative intensities of the radial beam array passing through the moderate turbulent atmosphere ($C_n^2 = 1 \times 10^{-15}$ m$^{-2/3}$) at different output planes: (a) $z = 0$, (b) $z = 250$ m, (c) $z = 1$ km and (d) $z = 5$ km and (e) $z = 10$ km, with other parameters $N = 30$, $m = 1$ and $a = 20w_0$.

In order to reveal the effect of the atmospheric turbulence on the propagation of optical vertex formed by using the coherent superposition of the radial beam array with the initial well-ordered phase distribution, figure 5 shows the evolution process of the normalized average intensity [$I(x,0,z)/I(x,0,z)_{max}$] of the radial beam array with different well-ordered phase distributions (i.e., different $m$), which leads to the formed optical vortices with different topological charges. From the above discussions in Fig. 3, we have known that in the moderate turbulent atmosphere, the optical vortex could be well formed at the output planes within the distance $z < 1$ km. In the Fig. 5(a), one can also see that the multi-annular central-dark optical vortices ($m > 0$) are well formed within this propagating distance $z < 1$ km. Note that there is no optical vortex for the case of $m = 0$. As $z$ increases, the optical vortex with topological charge $m = 1$ first completely disappears within the distance $5 \text{km} < z < 10 \text{km}$ [see Figs. 5(b) and 5(c)]; and then the optical vortex with $m = 3$ completely disappears within the distance $10 \text{km} < z < 50 \text{km}$ [see Figs. 5(c) and 5(d)]; and finally the optical vortex with $m = 5$ totally disappears within the distance $50 \text{km} < z < 100 \text{km}$ [see Figs. 5(d) and 5(e)]. Our theoretical results are very well in agreement with the recent theoretical results [36]. The atmospheric turbulence will strongly decreases the topological charge of the optical vortex and even destroys the optical vortex. The larger the topological charge, the longer distance the optical vortex could propagate over. Therefore using the optical vortex with the larger topological charge is useful for the long-distance atmospheric optical communication.

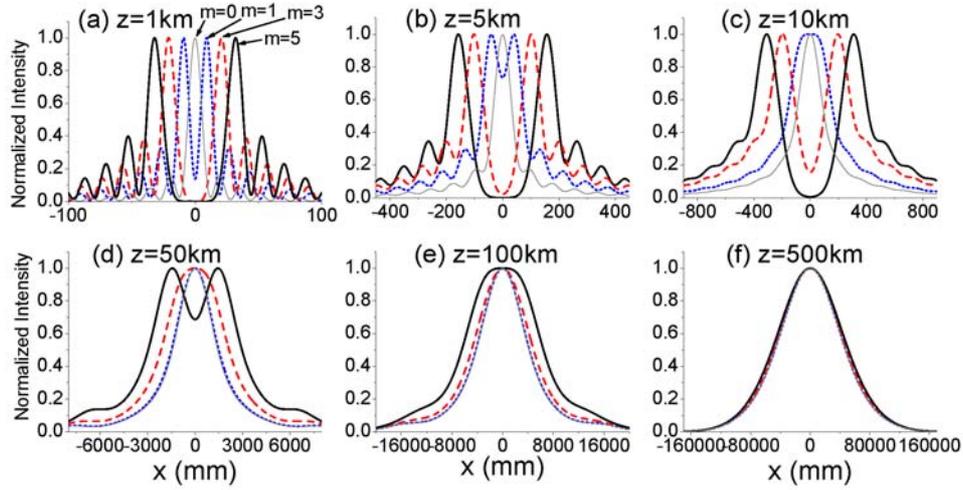

Fig. 5. (Color Online). Crossing lines ($y = 0$) of the normalized average intensity of the radial beam array propagating in the moderate turbulent atmosphere with different topological charges (thin gray curves: $m = 0$, dotted curves: $m = 1$, dashed curves: $m = 3$, and solid curves: $m = 5$) at different output planes: (a) $z = 1$ km, (b) $z = 5$ km, (c) $z = 10$ km and (d) $z = 50$ km, (e) $z = 100$ km, and (f) $z = 500$ km, with other parameters $N = 30$ and $a = 20 w_0$.

In Fig. 5, one can find that as the optical vortices disappear after the sufficient long-distance atmospheric propagation, the intensity profiles of the output beams of all the radial beam arrays with different initial phase distributions (i.e., different $m$) gradually become the same and tend to be the Gaussian shape [see Fig. 5(e) and 5(f)]. Note that the curve for $m = 1$ almost overlaps the curve for $m = 0$ in Fig. 5(d), and all curves are almost overlapped in Fig. 5(f). The reason is qualitatively given by the following: as the propagating distance $z$ increases, the atmospheric turbulence makes the coherent length of the light wave decrease, which first leads to the disappearance of the formed optical vortex or the prohibition of the formation of the optical vortex, depending on the strength of atmospheric turbulence; as the propagating distance z further increases, the turbulent atmosphere further leads to the decreasing of the light coherence and destroys the coherent superposition, which leads to the great broadening effect of the beam profiles (note the different scales in Fig. 5); after the sufficient long-distance propagation, the output beams actually become the *completely incoherent* light due to the effect of the turbulent atmosphere (i.e., the coherent length $\rho_0 \to 0$ as $z$ increases), which indicates that the output beams in the sufficient long-distance region have no memory on the information of the initial well-ordered phase distributions of the coherent radial beam arrays. Therefore the output beams have the same profiles and tend to the normal distributions (i.e., the Gaussian-shaped profiles) via the sufficient long-distance propagation.

**4. Summary**

We have investigated the effect of the atmospheric turbulence on the propagation of optical vertex formed by the radial coherent laser beam array. The propagation formula of such a radial coherent laser array, with the initial well-ordered phase distribution, passing through the turbulent atmosphere has been analytically derived from the generalized Huygens-Fresnel diffraction integral containing with the atmospheric factor. Based on the derived formula, the effect of the atmospheric turbulence on the propagation properties of the radial laser arrays has been analyzed in detail. It is found that in the case of the strong atmospheric turbulence, the strong decoherence effect of the strong turbulence on the light fields results in the prohibition for the formation of the optical vortices from the radial beam array (with the well-ordered phase distribution) even in the short-distance region and a rapid spreading of the output beam in the long-distance region; in the case of the moderate or weak atmospheric turbulence, the optical vortex could be first formed as similar to that in the free space in the short-distance region, however as the propagating distance increases, the formed optical vortices (the multi-annular centric-dark beams) gradually disappear due to the enhancement of the decoherence effect in the turbulent atmosphere. After the sufficient long-distance propagation, the output beam has no optical vortex and gradually becomes a Gaussian-shaped beam, and in this case the output beam is totally incoherent light due to the turbulent atmosphere.

**Acknowledgments**  This work was supported by the National Nature Science Foundation of China (10604047), by Zhejiang Province Scientific Research Foundation (G20630 and G80611) and by the financial support from RGC of HK Government (NSFC 05-06/01) and the financial support from Zhejiang Unviersity.